\documentclass{article}

\usepackage{cite}
\usepackage{graphicx,subcaption}
\usepackage{comment}
\usepackage{hyperref}

\makeatletter

\newcommand\autorefs[1]{\@first@ref#1,@}
\def\@throw@dot#1.#2@{#1}% discard everything after the dot
\def\@set@refname#1{%    % set \@refname to autoefname+s using \getrefbykeydefault
    \edef\@tmp{\getrefbykeydefault{#1}{anchor}{}}%
    \xdef\@tmp{\expandafter\@throw@dot\@tmp.@}%
    \ltx@IfUndefined{\@tmp autorefnameplural}%
         {\def\@refname{\@nameuse{\@tmp autorefname}s}}%
         {\def\@refname{\@nameuse{\@tmp autorefnameplural}}}%
}
\def\@first@ref#1,#2{%
  \ifx#2@\autoref{#1}\let\@nextref\@gobble% only one ref, revert to normal \autoref
  \else%
    \@set@refname{#1}%  set \@refname to autoref name
    \@refname~\ref{#1}% add autoefname and first reference
    \let\@nextref\@next@ref% push processing to \@next@ref
  \fi%
  \@nextref#2%
}
\def\@next@ref#1,#2{%
   \ifx#2@ and~\ref{#1}\let\@nextref\@gobble% at end: print and+\ref and stop
   \else, \ref{#1}% print  ,+\ref and continue
   \fi%
   \@nextref#2%
}
\makeatother

\usepackage{spconf,amsmath}
\usepackage{amssymb,bm}
\usepackage{mathtools}

\makeatletter
\newcommand{\specialcell}[1]{\ifmeasuring@#1\else\omit$\displaystyle#1$\fi}
\makeatother

\title{Graph Variogram: A novel tool to measure spatial stationarity}
\name{Alexander Serrano, Benjamin Girault, Antonio Ortega }
\address{Signal and Image Processing Institute, Department of Electrical Engineering,\\ University of Southern California}

\begin{document}

\setlength{\abovedisplayskip}{3pt}
\setlength{\belowdisplayskip}{3pt}

\ninept
\maketitle
\begin{abstract}
Irregularly sampling a spatially stationary random field does not yield a graph stationary signal in general. Based on this observation, we build a definition of graph stationarity based on intrinsic stationarity, a less restrictive definition of classical stationarity. We introduce the concept of graph variogram, a novel tool for measuring spatial intrinsic stationarity at local and global scales for  irregularly sampled signals by selecting subgraphs of local neighborhoods. Graph variograms are  extensions of variograms used for signals defined on continuous Euclidean space.  Our experiments with intrinsically stationary  signals sampled on a graph, demonstrate that graph variograms yield estimates with small bias of true theoretical models, while being robust to sampling variation of the space.
\end{abstract}


\section{Introduction}

Stationarity is a key characteristic used to describe random signals in both signal processing and geostatistical literature. In practice, the most common flavor of stationarity  is second order stationarity, which assumes both first and second order moments statistics exist and are invariant under arbitrary spatial displacements or time shifts \cite{HAYES}. Stationary signals are characterized in the spectral domain by uncorrelated frequencies, so that they can be analyzed and processed using the Power Spectrum Density (PSD) \cite{Papoulis.BOOK.1991}. These statistical assumptions allow for a wide range of applications where the signal measured is assumed stationary, such as ARMA models \cite{Box.BOOK.2015} or optimal filtering of noise (Wiener filters) \cite{Papoulis.BOOK.1991}. 

Geology, atmospheric science, ecology, among other disciplines that require data collection from different spatial locations, often use stationarity \cite{CRESSIE1}. 
In particular, spatial stationarity is appealing here as it conveys information about the random signal defined over a Euclidean space. 

Our objective is to build a definition of stationarity for signals defined on graphs, that features consistent descriptions of the random signal we observe, irrespective of how we sample continuous space.

Recent contributions to extending stationarity to graph signals have considered both global \cite{BG1,Girault.THESIS.2015,Perraudin.TSP.2017,MARQUES} and local \cite{BG5,BG4} definitions of graph stationarity. The former is summarized as second order graph stationarity characterized by uncorrelated spectral components leading to a straightforward definition of graph PSD \cite{Girault.THESIS.2015}. In the vertex domain, these signals are invariant through graph translation \cite{BG6,Girault.THESIS.2015}, hence directly extending the framework of second order stationarity for Euclidean domains using the graph translation instead of a spatial shift. Unfortunately, in this context of sampled Euclidean domains, relating the graph Fourier transform to the continuous Fourier transform is not an easy task, with the closest result being an asymptotic relation between graph Fourier modes and continuous Fourier modes when the number of sampled vertices grows to infinity  \cite{Belkin.NIPS.2006}. An important consequence of this observation is that the graph translation does not easily relate to a spatial shift, or diffusion operator, in the continuous domain.

To lift the difficulty of extending the shift operator for Euclidean domains to a sensible operator on graph signals, we propose to leave the definition of second order stationarity and work with \emph{intrinsic stationarity} of random fields defined over Euclidean domains. 

Intrinsic stationarity is tightly linked to the \textit{variogram}, and its empirical equivalent, which allows us to measure spatial correlations for different sized neighborhoods of nodes on both local and global scales. 
More precisely, instead of extending shifts and PSD, we propose 
a new measure of spatial stationarity for random sensor network signals defined over an irregular graph domain: The \textit{graph variogram}.

  Our proposed graph variogram extends the original variogram by combining graph structure with distance information between pairs of vertices. More generally, for a given graph structure, local graph variograms can be computed by selecting local subgraphs corresponding to local neighborhoods. Using these subgraphs, we are able to finely measure the local dependence of sample correlation with distance. 

  Finally, averaging the local dependence of sample correlation with distance over all these local subgraphs yields our graph variogram. 
  A major benefit of this definition is its robustness to 
  variation of the sampling used to measure the continuous signals of interest through the use of distances between samples: 
  the graph variogram provides consistent descriptions of spatial stationary signals, even with nonuniform samples.
 
 The rest of the paper is organized as follows. In \autoref{sec:background}, we review and compare definitions of stationarity from signal processing and geostatistics. We then describe the graph signal processing framework and graph stationarity in \autoref{sec:background_gsp}, and our definition of the graph variogram and its properties in \autoref{sec:graph_variogram}. Finally, \autoref{sec:experiments} illustrates our contribution on synthetic examples.                                 
 
\section{Background on Stationarity}
\label{sec:background}

\subsection{Second Order  Stationarity}
\label{sec:background:WSS}

For continuous spatial models, we denote the random field  defined at any particular spatial location $\mathbf{s} \in \mathbb{R}^{d}$ as $X(\mathbf{s})$. The collection of spatial data from observations of the random field at a subset of N locations $D \subset \mathbb{R}^{d}$ can be represented as a vector $\mathbf{x} \in \mathbb{R}^{N}$, where $x_i  = X(\mathbf{s_i}) , \forall \mathbf{s_i} \in D $. For sensor network applications we use either $D \subset \mathbb{R}^{2}$ ($d=2$) or $D \subset \mathbb{R}^{3}$ ($d=3$). 

In practice, second order or wide sense stationarity is an assumption used for modelling  spatial  datasets. Second order stationarity describes a random field whose first and second order moments exist and  are both invariant under spatial displacements $\forall \mathbf{s},\mathbf{h} \in \mathbb{R}^{d}$:
\begin{align}
  \mathbb{E}\big[X(\mathbf{s})\big] &= \mu\text{,} \label{eq:const_mean_second_order} \\ 
  C(\mathbf{s+h},\mathbf{s}) &= 
  \mathrm{cov}\big[ X(\mathbf{s+h}),X(\mathbf{s})\big] = C(\mathbf{h})\text{.} \label{eq:autocov}
\end{align} 

 $C(\cdot)$ in \eqref{eq:autocov} is called the autocovariance function.  A second order stationary random field whose autocovariance function $C(\mathbf{h})$ only depends on the magnitude of the displacement vector, $||\mathbf{h}|| = h$ is called \textit{isotropic}.

\subsection{Intrinsic  Stationarity}

The previous definition of stationarity is more commonly found in the signal processing literature. Instead, researchers in geostatistics focus on the first and second order moment  of the  difference $\smash{ X(\mathbf{s+h})-X(\mathbf{s})}$, rather than $\smash{X(\cdot)}$. Past work has shown that intrinsic stationarity is a less restrictive form of stationarity than second order stationarity \cite{MYERS}. An \textit{intrinsic} stationary random field \cite{CRESSIE1} satisfies the following  relations $\forall \mathbf{s},\mathbf{h} \in \mathbb{R}^{d}$:
\begin{align}
 \mathbb{E}\big[X(\mathbf{s+h}) - X(\mathbf{s})\big] &\ =   0   \label{eq:zero_mean} \\ 
 2\gamma(\mathbf{s+h},\mathbf{s}) &\coloneqq   \mathrm{var}\big[ X(\mathbf{s+h}) -X(\mathbf{s})\big]  = 2\gamma(\mathbf{h})
\end{align} 

where the quantity $2\gamma(\cdot)$ is known as the \textit{variogram}. $2\gamma(\cdot)$ measures the variance of the difference between two random field values at two corresponding locations separated by lag displacement $\mathbf{h}$. Moreover, if the variogram of the intrinsically stationary random field only depends on $||\mathbf{h}||  = h$, it is called isotropic.

An interesting connection can be made between the covariance function and the variogram for a second order stationary random field. 
Assuming \eqref{eq:const_mean_second_order} and \eqref{eq:autocov}, the variogram evaluated at spatial locations $\mathbf{s+h}$ and $\mathbf{s}$ verifies \cite{CRESSIE1}: 

\begin{align}
  2\gamma(\mathbf{s\!+\!h},\mathbf{s}) & = \mathbb{E}\big[ X(\mathbf{s\!+\!h})^{2}\big]+\mathbb{E}\big[ X(\mathbf{s})^{2}\big] - 2\mathbb{E}\big[ X(\mathbf{s\!+\!h})X(\mathbf{s})\big] \nonumber \\
    &= 2\big[ C(\mathbf{0}) -C(\mathbf{h}) \big], \label{eq:autocov_variogram}
\end{align}
which in fact proves that any second order stationary random field inherits intrinsic stationarity. The converse, however, is not verified, with Wiener--Levy processes as simple counter--examples  \cite{MYERS}.  

\subsection{Variogram \& Empirical Variogram}

The original variogram  for signals defined over continuous space was first introduced in 1963 \cite{MATH}. Letting $X(\mathbf{s})$ be the random field evaluated at a particular location $\mathbf{s}$ of a geometrical region $V$, one calculates the variogram for displacement $\mathbf{h}$ as follows:
\begin{align}
	2\gamma(\mathbf{h}) &=  \frac{1}{|V|}\iiint_{V} \big[X(\mathbf{s+h}) - X(\mathbf{s}) \big]^{2}\mathbf{ds}. \label{eq:matheron}
\end{align}
In order to calculate \eqref{eq:matheron}, all possible pairs of points differing by the exact displacement $\mathbf{h}$ would be sampled. 
Since it is impossible to sample at all pairs of spatial locations,  empirical variograms are computed instead using a finite, discrete set of spatial positions in Euclidean space. 

In practice, we are also limited to a finite number of realizations of the random field. An empirical measure of the variogram using data obtained at a discrete set of spatial positions $\{ \mathbf{s_i}\}_{i}$ under the constant mean assumption is the following empirical variance:
\begin{align}
 2 \hat{\gamma}(\mathbf{h})  =  \frac{1}{|\mathrm{N}(\mathbf{h})|}\sum_{(\mathbf{s_i},\mathbf{s_j}) \in \mathrm{N}(\mathbf{h})} \big( X(\mathbf{s_i}) - X(\mathbf{s_j}) \big)^{2}      \label{eq:empirical_variogram}
\end{align}
where $\mathrm{N}(\mathbf{h}) = \{ (\mathbf{s_i},\mathbf{s_j}) :  \mathbf{s_i} - \mathbf{s_j} \approx \mathbf{h} , \forall \mathbf{s_i},\mathbf{s_j} \in D\}$. When data are irregularly sampled from Euclidean space, the variogram is usually smoothed by using pairs with displacement vectors within some tolerance region described by $\mathbf{s_i - s_j } \approx \mathbf{h}$ \cite{CRESSIE1}.

\subsection{Empirical Local Variogram}

The empirical variogram shown in \eqref{eq:empirical_variogram} is used to measure spatial variations on a global scale for displacement $\smash{\mathbf{h}}$. Two additional separate hypotheses can be incorporated into an empirical variogram: isotropy and approximation using local neighborhoods \cite{CRESSIE2}. Under the isotropic assumption, let $\mathrm{N}(h)$ denote the set of pairs of spatial locations $ (\mathbf{s_i},\mathbf{s_j}) \in D $  such that $\mathbf{s_i}$ is approximately at distance $h$ from $\mathbf{s_j}$. For most applications in geostatistics, the variogram defined in \eqref{eq:matheron} is an increasing function of distance $h$, since for many applications, the farther both samples of the random field are from one another, the more they differ on average\cite{MATH}.

\begin{figure}[t]
\centering
\includegraphics[]{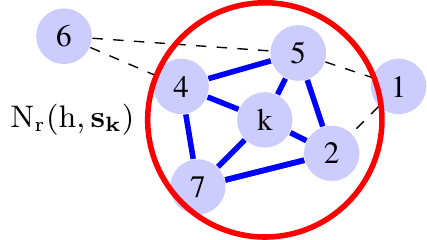}
\begin{comment}
\begin{tikzpicture}
  [scale=.34,auto=left] 
  \node [style={circle,fill=blue!20}] (n6) at (1,9.5) {6};
  \node [style={circle,fill=blue!20}] (n4) at (4.5,8)  {4};
  \node [style={circle,fill=blue!20}] (n5) at (8,9)  {5};
  \node [style={circle,fill=blue!20}] (n1) at (11,8) {1};
  \node [style={circle,fill=blue!20}] (n2) at (9,6)  {2};
  \node [style={circle,fill=blue!20}] (n3) at (7,7)  {k};
  \node [style={circle,fill=blue!20}] (n7) at (5,5)  {7};
  \node at (1.25,7) (Left)  {$\mathrm{N_{r}(h,\mathbf{s_k})}$} ;
  \foreach \from/\to in {n6/n4,n4/n5,n5/n1,n1/n2,n2/n5,n2/n3,n3/n4,n4/n7,n7/n2}
    \draw [dashed] (\from) -- (\to);
   \draw [dashed] (n6)  -- (n5);
   \draw [blue,ultra thick] (n3)  -- (n5);
   \draw [blue,ultra thick] (n3)  -- (n2);
   \draw [blue,ultra thick] (n3)  -- (n7);
   \draw [blue,ultra thick] (n3)  -- (n4);
   \draw [blue,ultra thick] (n5)  -- (n4);
   \draw [blue,ultra thick] (n7)  -- (n2);
   \draw [blue,ultra thick] (n7)  -- (n2);
   \draw [blue,ultra thick] (n2)  -- (n5);
   \draw [blue,ultra thick] (n7)  -- (n4);
   \draw [red, ultra thick] (7,7) circle [radius=3.5];;

\end{tikzpicture}
\end{comment}
    \vspace{-0.3cm}
	\caption{Illustration of $\mathrm{N}_{r}(h;\mathbf{s_k})$. Dashed lines indicate pairs of sensors that would be excluded in empirical isotropic local variogram calculations. Solid  lines indicate pairs of sensors that would be included. Circle defines window\lq s localization with respect to node $k$. \label{fig:neighborhood_example} }
\end{figure}

More precisely, we can study heterogeneous spatial variations using local neighborhoods. We define $\mathrm{N}_{r}(h;\mathbf{s_k}) = \{ (\mathbf{s_i},\mathbf{s_j}) \ | \ ||\mathbf{s_i} - \mathbf{s_j}|| \approx {h},   ||\mathbf{s_i} - \mathbf{s_k}|| \leq r ,||\mathbf{s_j} - \mathbf{s_k}|| \leq r , \forall \mathbf{s_i},\mathbf{s_j} \in D \}$.  \autoref{fig:neighborhood_example}  illustrates how $\mathrm{N}_{r}(h;\mathbf{s_k}) $ would look like for a sensor network modeled using a graph structure.

Using these two hypotheses, an empirical isotropic local variogram at spatial location $\mathbf{s_k}$ will have a similar form to that of shown in previous work \cite{CRESSIE2}: 
\begin{align}
 2 \hat{\gamma}(h;\mathbf{s_k})  =  \frac{1}{|\mathrm{N}_{r}(h;\mathbf{s_k}) |}\sum_{(\mathbf{s_i},\mathbf{s_j}) \in \mathrm{N}_{r}(h;\mathbf{s_k}) } \big[ X(\mathbf{s_i}) - X(\mathbf{s_j}) \big]^{2}   \label{eq:binary_local_variogram_empirical}  
\end{align}

The  empirical isotropic local variogram in \eqref{eq:binary_local_variogram_empirical} can be interpreted as a local unweighted average of squared differences between values of $X(\cdot)$ at $(\mathbf{s_i},\mathbf{s_j}) \in \mathrm{N}_{r}(h;\mathbf{s_k})$. A generalization of \eqref{eq:binary_local_variogram_empirical} using weighting function $\mathrm{W}(\mathbf{s_i},\mathbf{s_j};h;\mathbf{s_k})$ and a corresponding  normalization term $\mathrm{W}(h;\mathbf{s_k})$ is shown in \cite{CRESSIE2}:
\begin{align}
  \mathrm{W}(\mathbf{s_i},\mathbf{s_j};h;\mathbf{s_k})&= w_{h}(||\mathbf{s_i} - \mathbf{s_j}|| - h)w_{s_k}(||\mathbf{s_i} - \mathbf{s_k}||)  \nonumber \\ 
  &  \times  w_{s_k}(||\mathbf{s_j} - \mathbf{s_k}||) \label{eq:window_separable}\\
  \mathrm{W}(h;\mathbf{s_k}) &= \sum_{i,j}\mathrm{W}(\mathbf{s_i},\mathbf{s_j};h;\mathbf{s_k})  \label{eq:normalization_term}
  \end{align}
 where $w_h (\cdot)$ is a window that describes the tolerance region and $w_{s_k} (\cdot)$ defines the neighborhood of $\mathbf{s_k}$. This yields the generalized variogram:
  \begin{align}
 2 \hat{\gamma}(h;\mathbf{s_k})  &=  \sum_{ i,j} \frac{\mathrm{W}(\mathbf{s_i},\mathbf{s_j};h;\mathbf{s_k})}{\mathrm{W}(h;\mathbf{s_k}) }\big[ X(\mathbf{s_i}) - X(\mathbf{s_j}) \big]^{2}     \label{eq:local_variogram_empirical}
\end{align}
 The variogram formulation in  \eqref{eq:local_variogram_empirical} is equivalent to that of in \eqref{eq:binary_local_variogram_empirical} when $w_h (\cdot)$ and $w_{s_k} (\cdot)$ are binary valued (0-1) decreasing functions. 
 
\section{Background on Graph Signal Processing}
\label{sec:background_gsp}

\subsection{Fundamentals \& Notation}

Following the GSP literature \cite{SHU2,ortega2018graph}, we 
model sensor networks as undirected graphs $G(V,E)$ with a vertex set consisting of $N$ nodes. The adjacency matrix of the graph is denoted as $\mathbf{A} \in \mathbb{R}^{N\times N}$. If nodes $i$ and $j$ are connected through an edge weighted by $w_{ij}$, then $A_{ij}=A_{ji}=w_{ij}$. In the context of sensor networks, $i \in V$ represents the $i^{\text{th}}$ sensor, and $x_i$ denotes the value of the random field $X(\cdot)$ at the $i^{\text{th}}$ sensor's position $\mathbf{s_i}$. For all $N$ sensors, we define our graph signal as $\mathbf{x} \in \mathbb{R}^{N}$. The Euclidean distance between the $i^{\text{th}}$ and $j^{\text{th}}$ sensors is denoted as $d_{ij} = ||\mathbf{s_i}-\mathbf{s_j}||$.  \ \

There exist multiple ways to construct edge weights $w_{ij}$ to model local neighborhood relationships between data points \cite{SHU1}. We use here a fully connected graph by linking all nodes with edges weighted by a Gaussian kernel of the Euclidean pairwise distances $\smash{w_{ij} = \exp(-{d^{2}_{ij}}/{2\sigma^{2}} )}$. Alternatively, one can also reduce the number of edges by connecting the K-nearest neighbors to each vertex, with weights given by a Gaussian kernel. Given a fixed adjacency matrix $\mathbf{A}$, the combinatorial or unnormalized graph Laplacian matrix is defined as $\mathbf{L} = \mathbf{D} - \mathbf{A}$, where $\mathbf{D}$ is a diagonal degree matrix verifying $D_{ii} = \sum_{j}w_{ij}$. Conventional methods in graph signal processing  use the eigenvectors of $\mathbf{L}$ to define the graph Fourier transform (GFT) \cite{SHU2,ortega2018graph}.  

\subsection{Graph PSD and Graph Stationarity}
\begin{figure}[tb]

\centering

 \begin{subfigure}[b]{0.49\columnwidth}
 
    \centering
    
    \includegraphics[width=.6\linewidth,trim=85 35 45 30]{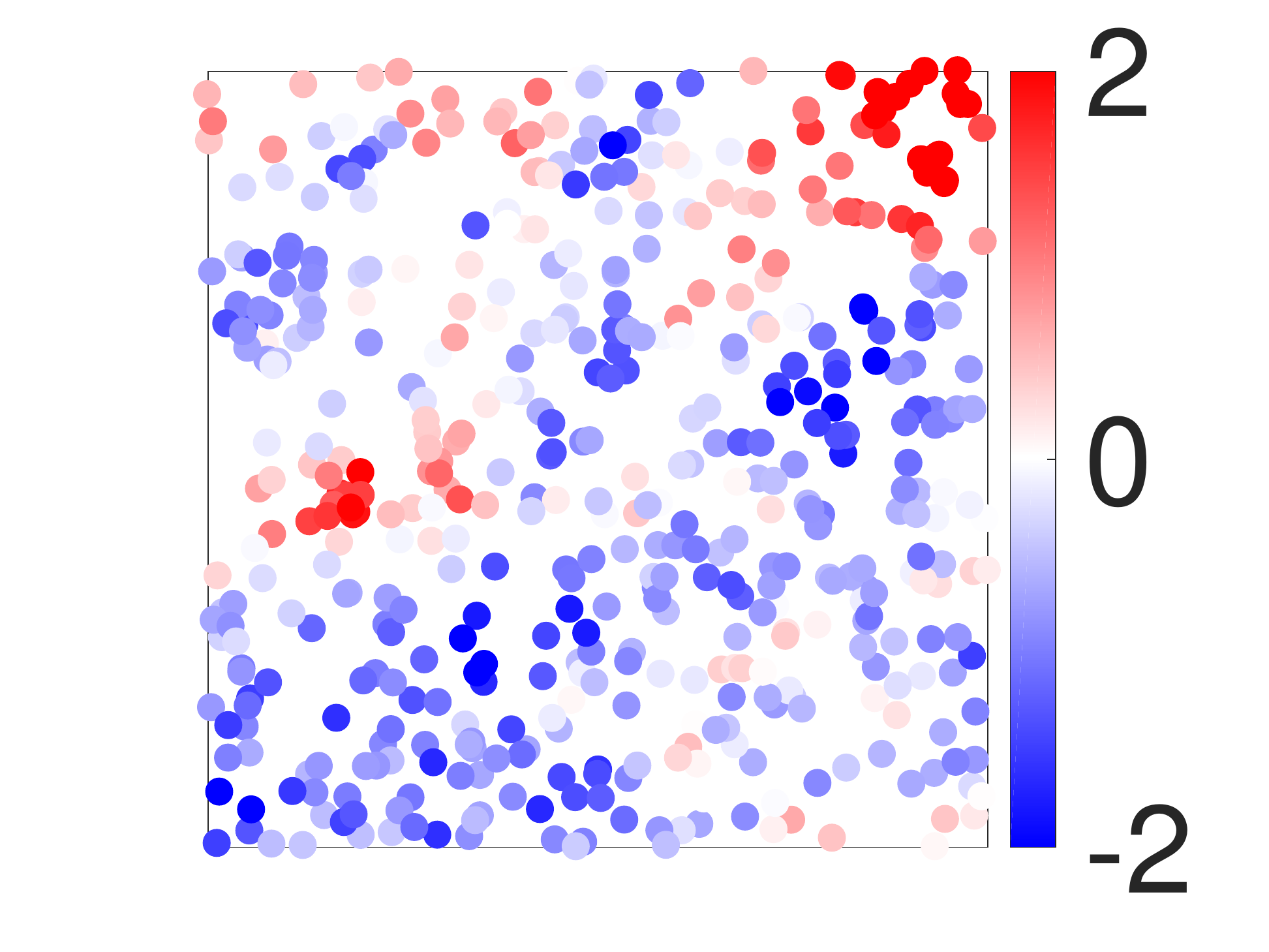}

    \vspace{-0.1cm}
    %% </TRICK>
    \caption{}
    \label{fig:results_PSD:uniform:Euclidean_sampling}
  \end{subfigure}
  \begin{subfigure}[b]{0.49\linewidth}
    \centering
    \includegraphics[width=.6\linewidth,trim=85 35 45 30]{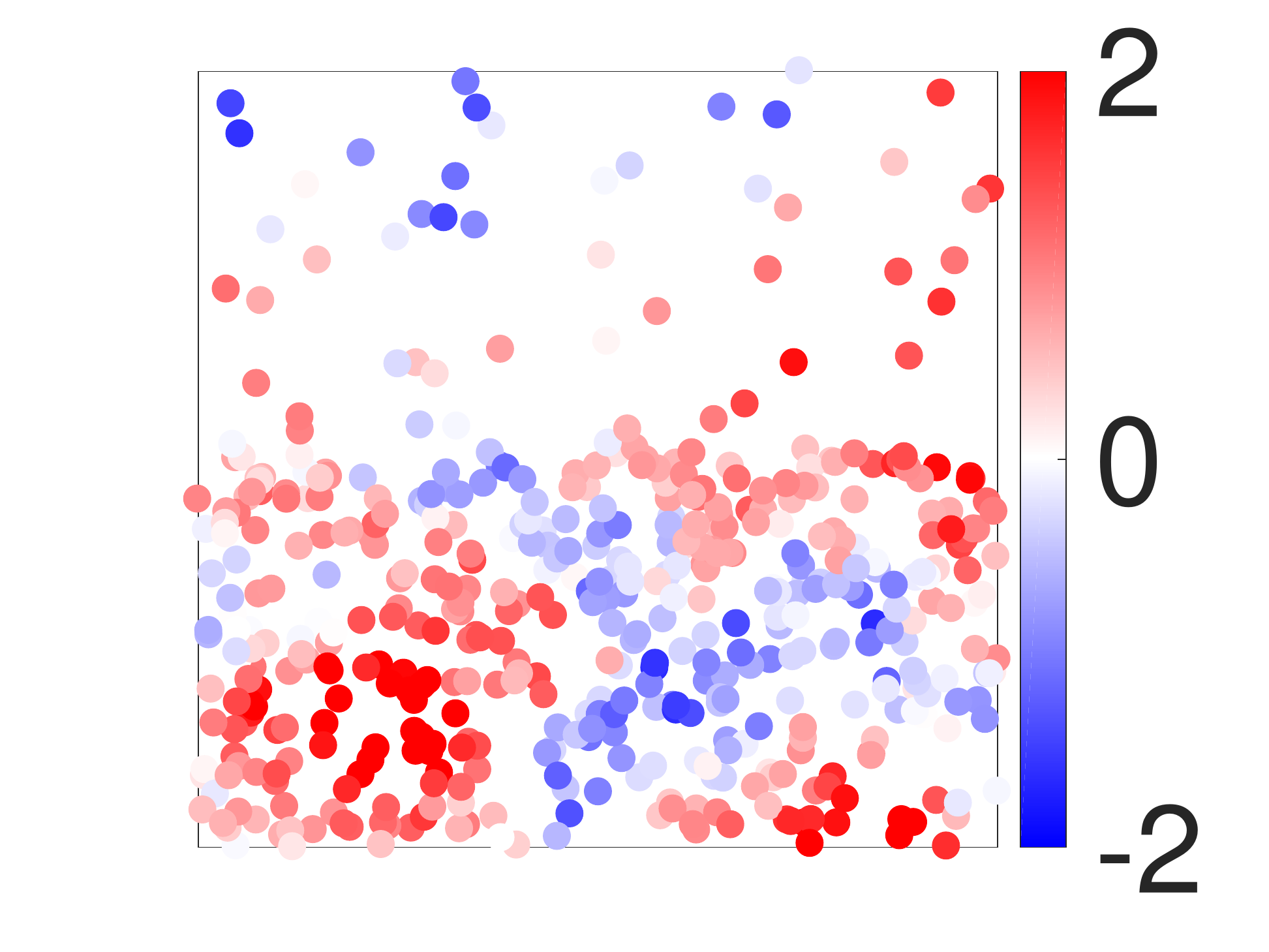}
    \vspace{-0.1cm}
    \caption{}
  \label{fig:results_PSD:nonuniform:Euclidean_sampling}
  \end{subfigure} 
%   \begin{subfigure}[b]{0.45\columnwidth}
%    \includegraphics[width=\linewidth]{uniform_log_spectral_autocorrelation.eps}
%    \subcaption{}
%    \label{fig:uniform:spectral_correlation}
%  \end{subfigure}
%  \begin{subfigure}[b]{0.45\columnwidth}
%    \includegraphics[width=\linewidth]{nonuniform_log_spectral_autocorrelation.eps}
%  \subcaption{}
%  \label{fig:nonuniform:spectral_correlation}
%  \end{subfigure} 

 \begin{subfigure}[b]{0.49\linewidth}
    \centering
   \includegraphics[]{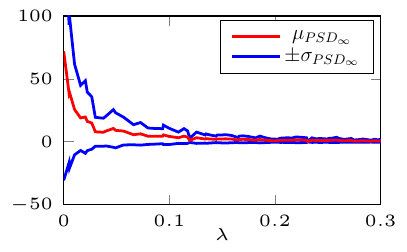}
  \begin{comment}
    \begin{tikzpicture}
      %\pgfplotstableread{U_data_100graphs.dat}\mytable
    %% <TRICK>
    % Lots of tricks in here... mostly to make the text smaller
      \begin{axis}[  	
    tiny,
  	width = 4.8cm,height=3.5cm,%
  	%xlabel={Graph Frequency, $\lambda$},
  	xlabel={$\lambda$},xlabel near ticks,xlabel shift=-0.2cm,x label style={font=\tiny},%
  	xmin=0, xmax =0.3,
  	xtick={0,0.1,0.2,0.3,0.4},%
  	ymin=-50, ymax=100,
    %ylabel={$PSD_{\infty}(\lambda)$},
    %legend entries={$\mu_{ PSD_{\infty}}\pm \sigma_{ PSD_{\infty}}$,$\mu_{  PSD_{\infty}      }$},
    %legend style={at={(0.5,-0.6)},anchor=north},
    %legend style={font=\tiny},
    legend style={nodes={scale=0.5, transform shape},font=\large},%
    reverse legend
    %legend pos=north east
    ]
    \pgfplotstableread{PSD_U_noknn_data_singlegraph.dat}{\UPSD}
    \addplot[mark = none,color = blue,  thick] table[x index=0,y index=2] {\UPSD};
    \addlegendentry{$\pm \sigma_{ PSD_{\infty}}$}
    \addplot[mark = none,color = blue,  thick,forget plot] table[x index=0,y index=3] {\UPSD};
    \addplot[mark = none,color = red,  thick] table[x index=0,y index=1] {\UPSD};
    \addlegendentry{$\mu_{ PSD_{\infty}}$}
      \end{axis}  
    %% </TRICK>
\end{tikzpicture}
\end{comment}
    \vspace{-0.1cm}
    \subcaption{}
    \label{fig:results_PSD:uniform:PSD}
  \end{subfigure}
  \begin{subfigure}[b]{0.49\columnwidth}
    \centering
    \includegraphics[]{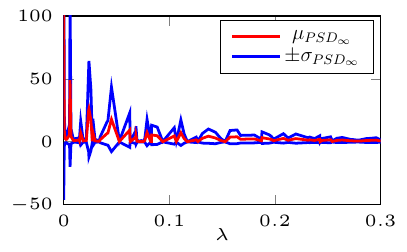}
	\begin{comment}
      \begin{tikzpicture}
      %\pgfplotstableread{U_data_100graphs.dat}\mytable
      \begin{axis}[  	
    tiny,
  	width = 4.8cm,height=3.5cm,%
  	%xlabel={Graph Frequency, $\lambda$},
  	xlabel={$\lambda$},xlabel near ticks,xlabel shift=-0.2cm,x label style={font=\tiny},%
  	xmin=0, xmax =0.3,
  	xtick={0,0.1,0.2,0.3,0.4},%
  	ymin=-50, ymax=100,
    legend entries={$\pm \sigma_{ PSD_{\infty}}$,$\mu_{ PSD_{\infty}}$},
     legend style={nodes={scale=0.5, transform shape},font=\large},%
    reverse legend
    ]
    \pgfplotstableread{PSD_NU_noknn_data_singlegraph.dat}{\NUPSD}
    \addplot[mark = none,color = blue,  thick] table[x index=0,y index=2] {\NUPSD};
     \addplot[mark = none,color = blue,  thick,forget plot] table[x index=0,y index=3] {\NUPSD};
     \addplot[mark = none,color = red,  thick] table[x index=0,y index=1] {\NUPSD};
  \end{axis}  
\end{tikzpicture}
\end{comment}
    \vspace{-0.1cm}
  \subcaption{}
  \label{fig:results_PSD:nonuniform:PSD}
  \end{subfigure} 
\vspace{-0.3cm}
\caption{Left column uniform sampling. Right column nonuniform sampling. (\subref{fig:results_PSD:uniform:Euclidean_sampling})-(\subref{fig:results_PSD:nonuniform:Euclidean_sampling}) Intrinsically stationary signal in Euclidean space with $\gamma(h) = 1 - \exp \big(-\frac{h}{0.2}\big)$ and sampled on the graph vertex locations. (\subref{fig:results_PSD:uniform:PSD})-(\subref{fig:results_PSD:nonuniform:PSD})  empirical graph PSD as a function of normalized graph frequency, $\lambda$. Note that for both sampling schemes high correlations for short distances ($\gamma(0) = 0$) lead to low energy in the higher spectrum since higher graph frequencies correspond to large variations on edges with large weights. }
\label{fig:results_PSD}
\end{figure}

 It is worthwhile to study how past contributions to graph stationarity using  the graph PSD describe intrinsic stationary signals. 
 Fully connected, uniformly and nonuniformly spaced graph structures are generated as shown in \autoref{fig:results_PSD:uniform:Euclidean_sampling} \autoref{fig:results_PSD:nonuniform:Euclidean_sampling} , respectively. For both sampling schemes, we compute spectral autocorrelation matrices and graph PSDs using 1000 graph signal realizations. Comparing \autoref{fig:results_PSD:uniform:PSD} with \autoref{fig:results_PSD:nonuniform:PSD}, we observe that even though the signals are spatially stationary, the graph PSD is far from being consistent across  realizations.

\section{Spatial Stationarity Using GSP}
\label{sec:graph_variogram}

%\pgfplotstableread{GTvariogram_data.dat}{\GTvariogram}

%In this section, we describe our approach extending the variogram to the graph setting leading to our definition of graph intrinsic stationarity.

\subsection{Relating the Variogram to the Graph Laplacian}

 In the context of sensor networks, we have direct access to pairwise distances between nodes, which allows us to introduce the concept of the variogram. 

For our problem formulation we use the \textit{graph Laplacian quadratic form} of a graph signal $\mathbf{x}$:
\begin{align}
    \mathbf{x^{T}Lx}  = \frac{1}{2}\sum_{i,j = 1}^{N} w_{ij}\big[ x_i - x_j \big]^{2}  = \sum_{\mathclap{(i,j) \in E}}  w_{ij}\big[ x_i - x_j \big]^{2},   \label{eq:quadform_original}
\end{align}
 where $\mathbf{x^{T}Lx}$ 
 %as defined in \eqref{eq:quadform_original}  
 measures squared differences of random field values across all edges $(i,j) \in E$. For constant equal  weights $w_{ij} = 1$, this quadratic form measures the variation of the graph signal on a global-scale.   

 Comparing \eqref{eq:local_variogram_empirical} and \eqref{eq:quadform_original}, we see that the empirical isotropic local variogram, $2 \hat{\gamma}(h;\mathbf{s_k})$, is a valid graph Laplacian quadratic form. More precisely, the normalized weights $\smash{\frac{\mathrm{W}(\mathbf{s_i},\mathbf{s_j};h;\mathbf{s_k})}{\mathrm{W}(h;\mathbf{s_k}) }}$ play the same role as that of $w_{ij}$ in \eqref{eq:quadform_original}. Computing the generalized empirical isotropic local  variogram is therefore equivalent to computing the quadratic form \eqref{eq:quadform_original} for a graph Laplacian matrix that best describes $\mathrm{N}_{r}(h;\mathbf{s_k})$.
 %Next we
 %We now have a simple alternative for computing empirical variograms on an irregular graph domain. The following section 
 %describe how to define the appropriate $\mathbf{L}$ for computing the empirical isotropic local variogram for $N_{r}(h;\mathbf{s_k})$. 

\subsection{Defining \texorpdfstring{$\mathrm{N}_{r}(h;\mathbf{s_k})$}{Nr(h;sk)} on a Graph}
%The graph Laplacian quadratic form method for computing $2 \hat{\gamma}(h;\mathbf{s_k})$ requires using a new Laplacian matrix with a set of weights that satisfy \eqref{eq:window_separable} where $w_h(\cdot)$ is accounted for.  
  Though they are similar, the key difference between \eqref{eq:local_variogram_empirical} and \eqref{eq:quadform_original} is that edge weights $w_{ij}$ in \eqref{eq:quadform_original} correspond to arbitrary distances. This motivates the need to bin  distances in order to make \eqref{eq:quadform_original} closely match the definition of the empirical isotropic variogram in \eqref{eq:local_variogram_empirical}. This amounts to creating a  different Laplacian for each $(h,\mathbf{s_k})$.

To make \eqref{eq:quadform_original} match more closely with \eqref{eq:local_variogram_empirical}, we first define a new adjacency matrix $\mathbf{{A}_{\boldsymbol{\Delta_h}}}$, that assigns zero weights to edges  with corresponding distance length outside prespecified bin width tolerance range ${\Delta_h} \coloneqq (h - \frac{\delta}{2},h +\frac{\delta}{2} )$. This  reduces to applying an elementwise binary operator on the original adjacency matrix.  
%Following the definition of  $\mathrm{N_r} (h,\mathbf{s_k})$used in \eqref{eq:binary_local_variogram_empirical}, we design a new adjacency matrix 
For any  pair of nodes $(i,j)$:  
\begin{align}
   [ \mathbf{A_{\boldsymbol{\Delta_h}}}]_{i,j}  & =   \left\{ \begin{array}{cc} 
                1, & \hspace{1mm} d_{ij} \in {\Delta_h} \\
                0, & \hspace{1mm}  d_{ij} \notin {\Delta_h}  \end{array} \right.  \label{eq:adjacency_previous}
\end{align}
The corresponding graph Laplacian $\mathbf{{L}_{\boldsymbol{\Delta_h}}}$ is calculated from $\mathbf{{A}_{\boldsymbol{\Delta_h}}}$. 

Suppose out of all pairs of nodes, the maximum distance is $d_{max}$. ${\Delta_h} $ is therefore contained within $(0,d_{max})$. We propose here to break the interval $(0,d_{max})$ into $H$ mutually disjoint intervals $\{ {\Delta_h} \}^{H}_{h=1}$  to obtain corresponding Laplacians $\{ \mathbf{{L}_{\boldsymbol{\Delta_h}}} \}^{H}_{h=1}$. We will study alternative binning methods in a future communication. By this stage, using the isotropic assumption we have built the necessary tools to compute the global variogram shown in \eqref{eq:empirical_variogram}.

 Alternatively, we can compute the generalized empirical local variogram shown in \eqref{eq:local_variogram_empirical}. To do so, we use non-binary windows centered on vertex $k$ decaying with distance to $k$. We propose using a graph signal $\mathbf{g_{k}} \in \mathbb{R}^{N}$ localized at  node $k$, to play the role of a local window in the vertex domain. Using $\mathbf{g_{k}}$ to modify our previous adjacency matrix in \eqref{eq:adjacency_previous}, we write  $\mathbf{{A}}_{\boldsymbol{(\Delta_h,k)}}$:
\begin{align}
\mathbf{{A}}_{\boldsymbol{(\Delta_h,k)}}  = \mathbf{G_{k}A_{\boldsymbol{\Delta_h}}}\mathbf{G_{k}}   &\quad \text{with} \quad&
 \mathbf{G_k} =  \mathrm{diag}\big( \mathbf{g_k} \big) 
  \end{align}
  
 The normalization constant that would be analogous to $W(h;\mathbf{s_k})$ in \eqref{eq:normalization_term} is accounted for by computing the quadratic form of the degree matrix $\mathbf{D}_{\boldsymbol{(\Delta_h,k)}}$:
 \begin{align}
     \mathbf{{W}}_{\boldsymbol{(\Delta_h,k)}} = \mathbf{1}^{T} \mathbf{{D}}_{\boldsymbol{(\Delta_h,k)}}\mathbf{1} \label{eq:degree_normalization}
  \end{align}
  Here  $\mathbf{D}_{\boldsymbol{(\Delta_h,k)}}$ accounts for degree variations as a function of both  $\Delta_h$ and $k$.
  
\subsection{Graph Variogram}

 We can now proceed with writing our final expression for calculating the GSP-based empirical isotropic local variogram. We call this quadratic form the graph local variogram $2\gamma_{G}$, where $2\gamma_{G}({\Delta_h},k)$ defines an empirical local variogram relative to center node $k \in V$ with respect to distance bin $\Delta_h$:
\begin{align}
   2\gamma_{G}({\Delta_h},k)= 2 \frac{\mathbf{x}^{T} \mathbf{{L}}_{\boldsymbol{(\Delta_h,k)}}\mathbf{x}}{\mathbf{1}^{T} \mathbf{{D}}_{\boldsymbol{(\Delta_h ,k)}}\mathbf{1}} \label{eq:local_graph_variogram}
\end{align}
where $ \mathbf{{L}}_{\boldsymbol{(\Delta_h , k)}} = \mathbf{{D}}_{\boldsymbol{(\Delta_h , k)}}- \mathbf{{A}}_{\boldsymbol{(\Delta_h , k)}}$. $2\gamma_{G}({\Delta_h},k)$ is undefined whenever the neighborhood described by ($\Delta_h ,\mathbf{s_k}$) is empty. The 2 in the r.h.s. of \eqref{eq:local_graph_variogram} is included to account for the quadratic Laplacian form in \eqref{eq:quadform_original} using each edge only once, whereas the original variogram in \eqref{eq:local_variogram_empirical} counts each edge twice.
We further  define the graph global variogram as the average over vertices of the graph local variogram:
\begin{align}
    2\gamma_{G}(\Delta_h) = \frac{1}{N}\sum_{k=1}^{N} 2\gamma_{G}(\Delta_h , k ) 
\end{align}
$2\gamma_{G}(\Delta_h)$ measures spatial variations on a global scale for given ${\Delta_h}$. Finally, we propose the following  definition of intrinsic global graph stationarity:

\textbf{Definition:} (Intrinsic Global Graph Stationarity) \textit{A signal $\mathbf{x}$ defined over G(V,E)  is intrinsically stationary if and only if }
\begin{align}
    \mathbb{E}\big[\gamma_{G}(\Delta_h ,k)\big] &= \mathbb{E}\big[ \gamma_{G}(\Delta_h)\big], \forall k \in V
\end{align}

Intuitively, this definition states that no matter the center vertex $k$, the local graph variogram is the same.% Therefore, this definition implements a shift invariance in the vertex domain that is closer to the classical definition of stationarity, as opposed to graph stationarity, which implements a shift invariance in the spectral domain \cite{BG1}.

\section{Results}
\label{sec:experiments}
To validate the definition of the global variogram, we now perform experiments showing closeness to the theoretical true variogram. More precisely, we compute realizations of an intrinsic stationary field with an isotropic true variogram 2$\gamma(h)$. Simultaneously, we generate an independent sampling of the field for each realization of the isotropic model with the goal of showing robustness to both sources of noise. Both uniform and non-uniform samplings (see  \autoref{fig:results_PSD:uniform:Euclidean_sampling} and \autoref{fig:results_PSD:nonuniform:Euclidean_sampling}) have been used within a square grid defined by $\{\forall (x,y) \in \mathbb{R}^{2}  : 0\leq x \leq1,  0\leq y \leq1   \}$
as shown in the following set of figures, which were generated using GraSP \cite{BG2}.
 
 We build the original graphs using two schemes: fully connected and sparse using $K$-nearest neighbors ($K$ =100). In both cases, the edge weights are chosen   
 using a Gaussian kernel of the Euclidean distance with parameter $\sigma = 0.05$.
\begin{figure}[tb]
\centering
    \begin{subfigure}[b]{0.49\columnwidth}
        \centering
        \includegraphics[]{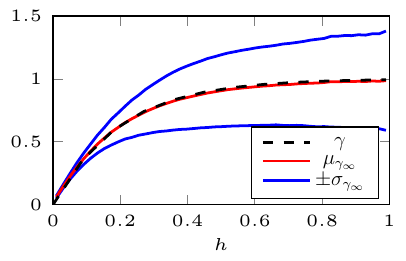}
        \begin{comment}
     \begin{tikzpicture}
      %\pgfplotstableread{U_data_100graphs.dat}\mytable
      \begin{axis}[  	
    tiny,
  	width = 5cm,height=3.5cm,%
  	%xlabel={Graph Frequency, $\lambda$},
  	xlabel={$h$},xlabel near ticks,xlabel shift=-0.1cm,x label style={font=\tiny},%
  	xmin=0, xmax =1,
  	xtick={0,0.2,...,1},%
  	ymin=0, ymax=1.5,
  	ytick={0,0.5,...,1.5},%
    %ylabel={$PSD_{\infty}(\lambda)$},
    %legend entries={$\mu_{ PSD_{\infty}}\pm \sigma_{ PSD_{\infty}}$,$\mu_{  PSD_{\infty}      }$},
    %legend style={at={(0.5,-0.6)},anchor=north},
    %legend style={font=\tiny},
    legend style={nodes={scale=0.5, transform shape},font=\large},
    legend entries={$\pm \sigma_{\gamma_{\infty}}$,$\mu_{\gamma_{\infty}}$,$\gamma$},
    legend pos=south east,%
    reverse legend
    ]
    \pgfplotstableread{U_NOKNN_data_singlegraph.dat}{\UNOKNNSINGLE}
    \addplot[mark = none,color = blue, thick] table[x index=0,y index=2] {\UNOKNNSINGLE};
     \addplot[mark = none,color = blue, thick,forget plot] table[x index=0,y index=3] {\UNOKNNSINGLE};
     \addplot[mark = none,color = red, thick] table[x index=0,y index=1] {\UNOKNNSINGLE};
     \addplot[mark = none,color = black, thick,dashed] table[x index=0,y index=1] {\GTvariogram};
  \end{axis}  
\end{tikzpicture}
\end{comment}
\vspace{-0.2cm}
\caption{}
\label{fig:uniform:noknn_singlegraph}
    \end{subfigure}
    \begin{subfigure}[b]{0.49\columnwidth}
        \centering
        \includegraphics[]{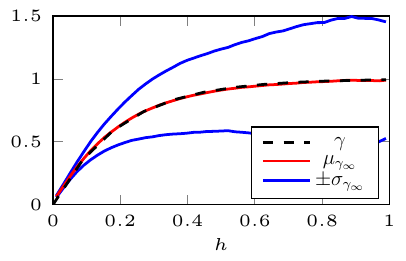}
	\begin{comment}
     \begin{tikzpicture}
      %\pgfplotstableread{U_data_100graphs.dat}\mytable
      \begin{axis}[  	
    tiny,
  	width = 5cm,height=3.5cm,%
  	%xlabel={Graph Frequency, $\lambda$},
  	xlabel={$h$},xlabel near ticks,xlabel shift=-0.1cm,x label style={font=\tiny},%
  	xmin=0, xmax =1,
  	xtick={0,0.2,...,1},%
  	ymin=0, ymax=1.5,
  	ytick={0,0.5,...,1.5},%
    %ylabel={$PSD_{\infty}(\lambda)$},
    %legend entries={$\mu_{ PSD_{\infty}}\pm \sigma_{ PSD_{\infty}}$,$\mu_{  PSD_{\infty}      }$},
    %legend style={at={(0.5,-0.6)},anchor=north},
    %legend style={font=\tiny},
    legend style={nodes={scale=0.5, transform shape},font=\large},
    legend entries={$\pm \sigma_{\gamma_{\infty}}$,$\mu_{\gamma_{\infty}}$,$\gamma$},
    legend pos=south east,%
    reverse legend
    ]
    \pgfplotstableread{NU_NOKNN_data_singlegraph.dat}{\NUNOKNNSINGLE}
    \addplot[mark = none,color = blue, thick] table[x index=0,y index=2] {\NUNOKNNSINGLE};
     \addplot[mark = none,color = blue, thick,forget plot] table[x index=0,y index=3] {\NUNOKNNSINGLE};
     \addplot[mark = none,color = red, thick] table[x index=0,y index=1] {\NUNOKNNSINGLE};
     \addplot[mark = none,color = black, thick,dashed] table[x index=0,y index=1] {\GTvariogram};
  \end{axis}  
\end{tikzpicture}
\end{comment}
\vspace{-0.2cm}
\caption{}
\label{fig:nonuniform:noknn_singlegraph}
    \end{subfigure} 
    
    \begin{subfigure}[b]{0.49\columnwidth}
        \centering
        \includegraphics[]{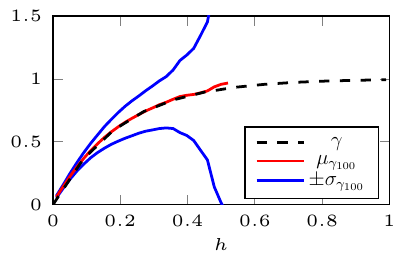}
        \begin{comment}
     \begin{tikzpicture}
      %\pgfplotstableread{U_data_100graphs.dat}\mytable
      \begin{axis}[  	
    tiny,
  	width = 5cm,height=3.5cm,%
  	%xlabel={Graph Frequency, $\lambda$},
  	xlabel={$h$},xlabel near ticks,xlabel shift=-0.1cm,x label style={font=\tiny},%
  	xmin=0, xmax =1,
  	xtick={0,0.2,...,1},%
  	ymin=0, ymax=1.5,
  	ytick={0,0.5,...,1.5},%
    %ylabel={$PSD_{\infty}(\lambda)$},
    %legend entries={$\mu_{ PSD_{\infty}}\pm \sigma_{ PSD_{\infty}}$,$\mu_{  PSD_{\infty}      }$},
    %legend style={at={(0.5,-0.6)},anchor=north},
    %legend style={font=\tiny},
    legend style={nodes={scale=0.5, transform shape},font=\large},
    legend entries={$\pm \sigma_{\gamma_{100}}$,$\mu_{\gamma_{100}}$,$\gamma$},
    legend pos=south east,%
    reverse legend
    ]
    \pgfplotstableread{U_KNN_data_singlegraph.dat}{\UKNNSINGLE}
    \addplot[mark = none,color = blue, thick] table[x index=0,y index=2] {\UKNNSINGLE};
     \addplot[mark = none,color = blue, thick,forget plot] table[x index=0,y index=3] {\UKNNSINGLE};
     \addplot[mark = none,color = red, thick] table[x index=0,y index=1] {\UKNNSINGLE};
     \addplot[mark = none,color = black, thick,dashed] table[x index=0,y index=1] {\GTvariogram};
  \end{axis}  
\end{tikzpicture}
\end{comment}
\vspace{-0.2cm}
 \caption{}
 \label{fig:uniform:knn_singlegraph}
    \end{subfigure}
     \begin{subfigure}[b]{0.49\columnwidth}
        \centering
       \includegraphics[]{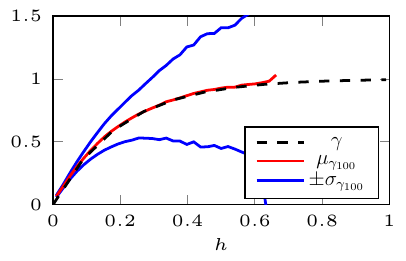} 
       \begin{comment} 
     \begin{tikzpicture}
      %\pgfplotstableread{U_data_100graphs.dat}\mytable
      \begin{axis}[  	
    tiny,
  	width = 5cm,height=3.5cm,%
  	%xlabel={Graph Frequency, $\lambda$},
  	xlabel={$h$},xlabel near ticks,xlabel shift=-0.1cm,x label style={font=\tiny},%
  	xmin=0, xmax =1,
  	xtick={0,0.2,...,1},%
  	ymin=0, ymax=1.5,
  	ytick={0,0.5,...,1.5},%
    %ylabel={$PSD_{\infty}(\lambda)$},
    %legend entries={$\mu_{ PSD_{\infty}}\pm \sigma_{ PSD_{\infty}}$,$\mu_{  PSD_{\infty}      }$},
    %legend style={at={(0.5,-0.6)},anchor=north},
    %legend style={font=\tiny},
    legend style={nodes={scale=0.5, transform shape},font=\large},
    legend entries={$\pm \sigma_{\gamma_{100}}$,$\mu_{\gamma_{100}}$,$\gamma$},
    legend pos=south east,%
    reverse legend
    ]
    \pgfplotstableread{NU_KNN_data_singlegraph.dat}{\NUKNNSINGLE}
    \addplot[mark = none,color = blue, thick] table[x index=0,y index=2] {\NUKNNSINGLE};
     \addplot[mark = none,color = blue, thick,forget plot] table[x index=0,y index=3] {\NUKNNSINGLE};
     \addplot[mark = none,color = red, thick] table[x index=0,y index=1] {\NUKNNSINGLE};
     \addplot[mark = none,color = black, thick,dashed] table[x index=0,y index=1] {\GTvariogram};
   
  \end{axis}  
\end{tikzpicture}
\end{comment}
\vspace{-0.2cm}
\caption{}
\label{fig:nonuniform:knn_singlegraph}
    \end{subfigure} 
%legend entries={$\pm \sigma_{\langle\gamma_\infty(h)\rangle_G}$,$\mu_{\langle\gamma_\infty(h)\rangle_G}$,$\gamma$}
  %%%%%%%%
     \begin{subfigure}[b]{0.49\columnwidth}
        \centering
		\includegraphics[]{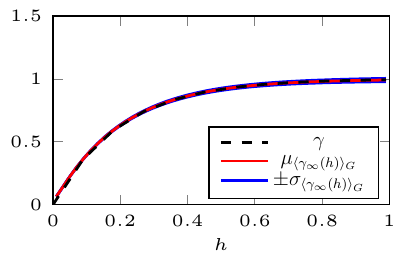}
\vspace{-0.2cm}
    \caption{}
    \label{fig:uniform:noknn_100graphs}
    \end{subfigure}
     \begin{subfigure}[b]{0.49\columnwidth}
        \centering
        \includegraphics[]{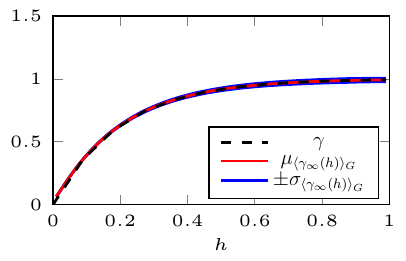}
     \begin{comment}
     \begin{tikzpicture}
      %\pgfplotstableread{U_data_100graphs.dat}\mytable
      \begin{axis}[  	
    tiny,
  	width = 5cm,height=3.5cm,%
  	%xlabel={Graph Frequency, $\lambda$},
  	xlabel={$h$},xlabel near ticks,xlabel shift=-0.1cm,x label style={font=\tiny},%
  	xmin=0, xmax =1,
  	xtick={0,0.2,...,1},%
  	ymin=0, ymax=1.5,
  	ytick={0,0.5,...,1.5},%
    legend style={nodes={scale=0.5, transform shape},font=\large},
    legend entries={$\pm \sigma_{\langle\gamma_\infty(h)\rangle_G}$,$\mu_{\langle\gamma_\infty(h)\rangle_G}$,$\gamma$},
    legend pos=south east,%
    reverse legend
    ]
    \pgfplotstableread{NU_data_100graphs.dat}{\NUDataHundred}
    \addplot[mark = none,color = blue, thick] table[x index=0,y index=2] {\NUDataHundred};
     \addplot[mark = none,color = blue, thick,forget plot] table[x index=0,y index=3] {\NUDataHundred};
     \addplot[mark = none,color = red, thick] table[x index=0,y index=1] {\NUDataHundred};
     \addplot[mark = none,color = black, thick,dashed] table[x index=0,y index=1] {\GTvariogram};

  \end{axis}  
\end{tikzpicture}
\end{comment}
\vspace{-0.2cm}
    \caption{}
    \label{fig:nonuniform:noknn_100graphs}
    \end{subfigure} 
\vspace{-0.3cm}
   \caption{Left column uniform sampling. Right column nonuniform sampling. (\subref{fig:uniform:noknn_singlegraph})-(\subref{fig:nonuniform:noknn_singlegraph}) Statistics of global variogram using 1 fully connected graph and 1000 signal realizations according to $\gamma(h) =  1 - \exp \big(-\frac{h}{0.2}\big)$. (\subref{fig:uniform:knn_singlegraph})-(\subref{fig:nonuniform:knn_singlegraph}) Statistics of global variogram using a single 100-nearest neighbor graph and 1000 signal realizations. There is a shorter support for the global variogram in the uniform sampling scheme since KNN removes most samples across longer distance ranges for  uniform structures than it does for the nonuniform structure. (\subref{fig:uniform:noknn_100graphs})-(\subref{fig:nonuniform:noknn_100graphs}) Statistics of global variograms over 100 fully connected graphs each with 1000 signal realizations.  }
   \label{fig:results_VARIOGRAM}
\end{figure}

Each of the four cases allows us to compute covariance matrices $\boldsymbol{\Sigma_x}$ for our graph signals defined at $N = 500$ spatial locations. Using these covariance matrices, we generate 1000 realizations of isotropic intrinsically stationary graph signals as $\mathbf{x} \sim \mathcal{N}(\mathbf{0},\boldsymbol{\Sigma_x})$ and compute empirical averages  and standard deviations of the empirical global graph variogram. In this section, we are interested in global graph variograms and therefore use $\mathbf{g_k = 1}$. We leave the study of local variograms and different $\mathbf{g_k}$ for future work.

From \autoref{fig:results_VARIOGRAM} we observe higher variance in empirical global variogram measures for edge sets $\Delta_j$ corresponding to longer distance intervals $\Delta_h$. Due to the fact that the spatial range of our nodes is limited, $|\mathrm{N}\big(h;\mathbf{s_k}\big)|$ decreases with increasing $h$. However, comparing and contrasting \autoref{fig:uniform:noknn_singlegraph} with \autoref{fig:nonuniform:noknn_singlegraph} as well as \autoref{fig:uniform:knn_singlegraph} with \autoref{fig:nonuniform:knn_singlegraph}, we obtain estimates of the true semivariogram model $\gamma(h)$ with small bias, regardless of the node spatial distribution.  \autoref{fig:uniform:noknn_100graphs} and \autoref{fig:nonuniform:noknn_100graphs}  both show low variance and bias  of global variograms across 100 realizations of graph structures. 

\section{Conclusions}

This work extends the concept of the variogram to stochastic signals defined over Euclidean spaces. 
Experiments involving intrinsically globally stationary theoretical semivariogram models suggest that the global variogram renders consistent statistical measures irrespective of uniform/nonuniform spatial graph structure. Overall, this contribution best served to provide an alternative perspective on graph stationarity for a new class of stochastic graph signals and allows for a definition of shift invariance in the vertex domain that matches that of the Euclidean domain. Future work will include more variogram models, more complex windowing operations, and distance binnings. Finally, we are looking
into defining a local stationarity test according to the local graph variogram definition as well as performing variogram analysis using graph filters.

%\vfill
%\pagebreak

\label{sec:refs}

\clearpage
\newpage
\bibliographystyle{IEEEbib}
\bibliography{refs}

\end{document}